\documentclass[a4paper,twoside]{article}
\newcommand{\affil}[1]{$^{\rm #1}$}
\usepackage[authoryear]{natbib}
\bibpunct{(}{)}{;}{a}{}{,}
\usepackage{graphicx}
\usepackage{color}
\usepackage{wasysym}
\usepackage{verbatim}
\usepackage{gensymb}	
\date{} 
\title{\large\bf\flushleft Fast algorithms for matching CCD images to a stellar catalogue}
\author{\parbox{\textwidth}{\flushleft
\vspace{-0.5cm}
{\it V. Tabur\affil{A,B}}\\
\vspace{0.4cm}
{\small \affil{A}\,School of Physics A28, University of Sydney, NSW 2006, Australia}\\
{\small \affil{B}\,Email: tabur@physics.usyd.edu.au}}}
\begin{document}
\twocolumn[
\begin{changemargin}{.8cm}{.5cm}
\begin{minipage}{.9\textwidth}
\vspace{-1cm}
\maketitle
%
%
\small{\bf Abstract:}

Two new algorithms are described for matching two dimensional coordinate lists of point sources that are significantly faster than previous methods. By matching rarely occurring triangles (or more complex shapes) in the two lists, and by ordering searches by decreasing probability of success, it is demonstrated that very few candidates need be considered to find a successful match. Moreover, by immediately testing the suitability of a potential match using an efficient mechanism, the need to process the entire candidate set is avoided, yielding considerable performance improvements. Triangles are described by a cosine metric that reduces the density of triangle space, permitting efficient searches. An alternative shape characterization method that reduces computational overhead in the construction phase is discussed. The algorithms are tested on a set of 10 063 wide-field survey images, with fields-of-view up to 4.8\degree x 3.6\degree, successfully matching 100\% of the images in a mean elapsed time of 6 ms (2.4 GHz Athlon CPU). The elapsed time of the searching phase is shown to vary by less than 1 ms for list sizes between 10 and 200 points, demonstrating that fast, robust searches may be completed in nearly constant time, independent of list size.


\medskip
\medskip
\end{minipage}
\end{changemargin}
]
\small

\section{Introduction}

In the course of carrying out a wide-field CCD imaging survey, two new methods for correlating the images to star catalogues have been developed, motivated by the need to efficiently handle the large number of stellar sources present on the images. Most previously published algorithms successfully cater for small lists ($\le$ 50 stars), but do not scale well to wide-fields containing $10^3$ or more stellar sources.

The problem of matching coordinate lists of point sources is a necessary prerequisite for deriving an astrometric plate solution. The objective is to match a subset of stars found on an image to their corresponding entries in a stellar catalogue in order to determine the transformation between detector coordinates and sky coordinates. The algorithm must handle translation and rotation, and small changes in scale caused by temperature related changes in focal length. In addition, it must cope with additional and missing stars. That is, the two lists may only partially overlap.

The efficiency of the algorithm is of paramount concern, since it is embodied within the closed-loop pointing system of the telescope and therefore affects the duty-cycle time, and ultimately constrains the number of images that can be acquired each night. Surveys that require very high photometric precision typically seek to accurately align their fields on the same detector pixels each night to overcome residual flat-fielding errors \citep{Everett}, and would benefit from the efficiency gains of a fast matching algorithm. Similarly, high cadence surveys, such as the Southern Sky Survey \citep{Keller} could improve precision and reduce its duty-cycle by utilizing a fast closed-loop pointing algorithm. Moreover, real-time attitude adjustments on spacecraft might be possible with the aid of an efficient matching algorithm to analyze on-board star camera images \citep[see for example][]{Fraser}. 

A number of algorithms have been proposed to solve this problem. \citet{Groth} describes an algorithm that matches geometrically similar shapes (triangles) in the two lists. By limiting the number of triangles constructed, and by only matching those triangles whose ratio of longest to shortest side are within a defined limit, his matching phase has a computational complexity of $O(n^{4.5})$ where $n$ is the number of stars in each list. \cite{Stetson} describes a very similar algorithm that he developed independently at around the same time. 

\citet{Murtagh} reviews a number of approaches and proposes his own, based upon characterization of a set of coordinates couples, with matching based on the proximity of feature vectors in the two lists. His method's matching phase has a computational complexity of $O(n^2)$. 

Nevertheless, Groth's algorithm appears to be the most widely accepted, with the methods applied across disciplines. For example, \citet{Arz} discuss its application to the problem of computer-aided identification of whale sharks, while \citet{Mars}, building upon the work of \citet{Groth}, describe an optimization to the voting phase of the algorithm, concluding that their method reduces the need for complicated filtering methods while successfully reducing the number of false matches.

More recently, \citet{PB} describe another variation of triangle matching, optimized to handle large lists of objects extracted from wide field images. Large fields contain thousands of stars and pose a severe test for matching algorithms, requiring efficient methods to accommodate the large number of point sources.

The following sections discuss two new methods for pattern matching that have a matching phase with a complexity that is nearly $O(1)$, at the cost of a slight loss in generality. They are collectively referred to as \emph{Optimistic Pattern Matching (OPM)} because they assume that (i) a good match is likely to be found, and (ii) the scale of the image is approximately known, thus permitting the use of an early exit strategy whereby only a small percentage of the candidate list is examined. By contrast, previous methods assumed an unknown scale which required the entire candidate list to be processed to determine the most likely match using a statistical approach. This required additional phases and complexity. In practice, an \emph{a priori} knowledge of an instrument's focal length is common place, and the use of a more general algorithm that assumes it is unknown mandates strategies that unnecessarily degrade performance.

Section 2 describes the algorithms in detail. $OPM_A$ is based upon a new definition of triangle space, while $OPM_B$ uses an alternative shape characterization method. Section 3 tests their performance using a large sample of survey images and compares them to earlier methods. Conclusions are summarized in Section 4.

\section{Algorithms}
The $OPM$ algorithm has some similarity to previous algorithms in that it attempts to match triangles in the two lists. However, it differs fundamentally by searching for rarely occurring triangles that are unique (or nearly so) to the field. By ordering the triangles by their estimated selectivity, and by testing the rarest shapes first, a correct match in usually identified extremely quickly. Thus, only a small fraction of the candidate list must be searched, allowing the search process to terminate early.

\subsection{List Creation}

The image for which a transformation is to be derived is first processed by a stellar detection routine to construct a list of sources ordered by descending magnitude. Each star is assigned an approximate instrumental magnitude estimated from the (non-sky subtracted) signal contained within the pixels attributed to the star. By assuming a uniform sky background and ignoring the effects of partial pixels, the method is computationally efficient in deriving an estimate of the relative intensity of the stellar sources found on the image. The brightest $n$ stars are selected from the list to form the image star list, denoted as $\mathcal{I}$.
The approximate equatorial coordinates of the field center are retrieved from the image header, together with the approximate focal length of the optics and the detector's physical dimensions, allowing the field of view (FOV) to be estimated. Using these quantities, the \emph{Hubble Guide Star} catalogue \citep{Lasker} is read to extract a list of the $n$ brightest catalogue stars within the field. This list of reference stars is denoted by $\mathcal{R}$. 

The $n$ brightest stars from each list are selected with the expectation that most will have a corresponding entry in the other list. However, experience shows that not all $\mathcal{I}$ will have a corresponding match in $\mathcal{R}$. Some uncertainty in the field center and, more importantly, differences in the passbands of the detector and catalogue results in different stars being selected. Increasing the size of $\mathcal{R}$ increases the probability that more $\mathcal{I}$ will be matched, at the expense of a longer triangle construction phase. Unlike previous methods, increasing list sizes does not adversely affect $OPM$'s matching performance in any significant way. It must be emphasized that only 3 stars common to both lists are necessary in order to find a successful match, but increasing $n$ increases the chance of an unusually shaped triangle being formed, which facilitates an early exit from the matching phase.

\subsection{$OPM_A$}

\subsubsection{Triangle Construction}

\begin{figure}[h]
\begin{center}
\includegraphics[scale=0.4, angle=0]{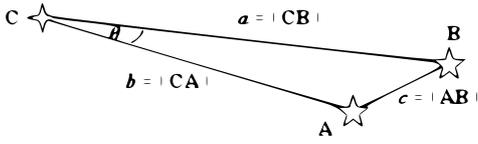}
\caption{$OPM_A$ nomenclature.}\label{DPfigure}
\end{center}
\end{figure}

Triangles are constructed from the stars in both lists. Each set of 3 stars (triplet) may be matched in 6 different ways with a triplet from the other list. Using an optimization introduced by \citet{Groth}, the number of candidates is reduced by a factor of 6 by assigning the vertices of the triangle such that vertices A and B define the shortest side, B and C the longest side, and A and C define the intermediate length side (see Figure \ref{DPfigure}). This scheme generates 
\begin{equation}\label{eq:1}
T =  n (n - 1) (n - 2) / 6
\end{equation}
unique triangles ($T$)  from a list of $n$ points. Next, we wish to assign some metrics to each triangle to describe its properties. \citet{Groth} used the ratio of the longest to the shortest side and the cosine of the angle at vertex B to define its position in a two-dimensional triangle space. \citet{Valdes} used the ratios of two sides, ($\frac{b}{a}, \frac{c}{a}$) where $a$, $b$ and $c$ are the side lengths in decreasing order, to define its location in triangle space. \citet{PB} defined a more elaborate scheme based on the side lengths and some auxiliary quantities. Although more computationally complex, their definition preserves chirality and maps triangles using a continuous function, cleverly avoiding discontinuities where small measurement errors may result in triangles being mapped to different parts of triangle space.

The $OPM_A$ algorithm defines triangle space as $(x_t, y_t)$, where 
\begin{equation}\label{eq:2}
x_t = \vec{CB} \cdot \vec{CA},\; y_t = \frac{a}{c}
\end{equation}
with $\vec{CB}$ and $\vec{CA}$ being the vectors from vertex C to B, and C to A respectively, and $a/c$ is the ratio of the length of the longest to the shortest side (Figure \ref{DPfigure}).

A dot product, or \emph{cosine metric}, is commonly used in text-based matching applications to compare the similarity of strings \citep[see for example][]{Rawat}. It has a number of useful properties, being stable under translation and rotation, and is computationally efficient to calculate using the relation: 
\begin{equation}\label{eq:3}
\vec{X} \cdot \vec{Y}  =  |\vec{X}| |\vec{Y}| \cos \theta  = \sum_{i} x_i y_i.
\end{equation}

However, its primary advantage over the other representations is that it provides a scalar value that is a function of the lengths of the two vectors and the angle between them. Therefore, it is useful in discriminating between the set of triangles that share the same side length ratios, but with different perimeters. Such triangles map to the same location in triangle space when using the definition of \citet{Groth} or \citet{Valdes}, requiring additional algorithmic complexity to separate the false matches that they produce.  

$OPM$ triangle space is sparse compared to that of \citet{Valdes}, who compressed all triangles into the range 
(0 \textless $x_t \le 1$, 0 \textless $y_t \le 1$), and \citet{PB} who used a domain of (-1 \textless $x_t$ \textless 1, -1 \textless $y_t$ \textless 1). \citet{Groth} used a cosine of one of the angles, restricting $0 \le x_t \le 1$, and arbitrarily constrained $y_t \le 10$. $OPM's$ definition permits an unconstrained range of values, thereby lowering the density (points per unit area) of triangle space, thus reducing the probability of misidentification. 

\begin{figure}[h]
\begin{center}
\includegraphics[scale=0.6, angle=0]{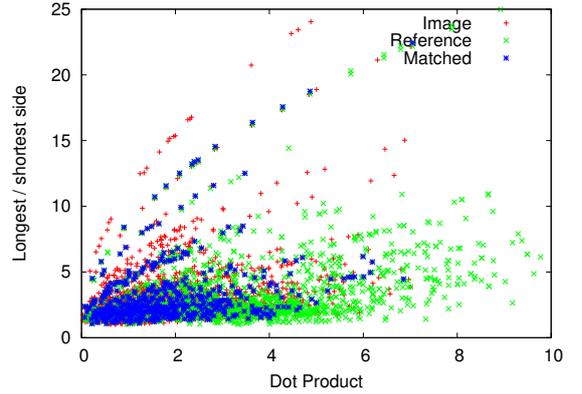}
\caption{$OPM_A$ triangle space.}\label{DPtrianglespace}
\end{center}
\end{figure}

Figure \ref{DPtrianglespace} plots $OPM_A$ triangle space for a representative image. Triangles formed from $\mathcal{I}$ and $\mathcal{R}$ are plotted using red pluses and green crosses respectively. A value of $n$ = 25 was used, resulting in 2300 triangles in each list. Two interesting features are immediately apparent. Firstly, the vast majority of triangles occur near the origin of the plot, where the density of points is greatest. Searches conducted in this region are very expensive due to the large number of candidates that must be considered. Secondly, a number of curving rows emanating from the origin and reaching up to large values of $x_t$, and/or $y_t$ are visible.

Each curve represents the set of triangles formed by a close pair of stars with a third more distant one. As the distance to the third star increases, the lengths of the two longest sides increase and the angle at vertex C becomes more acute, resulting in larger dot product. Similarly, the ratio of the longest to the shortest side increases. A key feature is that these curving rows are rather distinct, with the points furthest from the origin having very few neighbors. Processing the outlying points is very cost-effective due to the low number of candidates that must be considered.

The plot also shows the $\mathcal{I}$ / $\mathcal{R}$ pairings that were verified to be correct (blue pluses). Obviously, a few rows of image stars have no analogue extracted from the catalogue. This was caused by differences in relative magnitude of the stars in the two lists, due primarily to passband disparities. Similarly, some $\mathcal{R}$ have no matching $\mathcal{I}$ for the same reason. As expected, increasing the size of the $\mathcal{R}$ list (to 55 in this case) results in matches for all $\mathcal{I}$.

Triangle construction has a computational complexity of $O(n^3)$. However, by implementing an optimization proposed by \citet{Valdes}, that avoids calculating the same side length multiple times, the number of length calculations has been reduced from $\sim T^3$ to $\sim T^2$, with a proportional decrease in elapsed time. 

\subsubsection{Matching Triangles}

Searching for matching triangles in triangle space is a combinatorial problem. In principle, all triangles generated from $\mathcal{I}$ and $\mathcal{R}$ lists must be compared. A match is deemed to occur when a point in $\mathcal{I}$ triangle space is found to be within a certain tolerance $\epsilon$ of a point in $\mathcal{R}$ triangle space. 

A brute force method that compares each triplet of $\mathcal{I}$ stars to the entire list of $\mathcal{R}$ triplets is an expensive operation of $O(n^6)$. However, by sorting the $\mathcal{R}$ triangles by $y_t$ and using a binary search to find the starting point within the list, a large number of comparisons may be avoided. Only the points falling within $y_t \pm \epsilon$ need be compared. The choice of limiting searches using $y_t$ instead of $x_t$ is important, since it minimizes the number of candidates that fall within $y_t \pm \epsilon$, particularly when $y_t$ is large. The values in each coordinate are compared and a match is declared when they are within 2\%, the tolerance having been determined empirically from test data.

\subsubsection{Early-Exit Strategy}

The $OPM_A$ definition of triangle space ensures that triangles formed by two close vertices and a third more distant vertex map to sparse regions in triangle space, far from the densest areas occupied by triangles with similar side lengths. This property is exploited by searching the lowest density regions first, in the hope that a match will be found very quickly, allowing the process to terminate before the higher density areas must be considered. Each $\mathcal{I}$ triangle is assigned a score defined as the product of $x_t$ and $y_t$, and the list is sorted into descending order of score. Processing triangles in this order ensures that the relatively rare (highly selective) triangles are matched first. Comparisons are inexpensive, because there are few similar $\mathcal{R}$ candidates, and the few candidates that are within range are likely to be true matches.

\subsubsection{Checking a Match}

All potential matches require verification, since false matches will always be present. \emph{Voting} \citep{Groth,Valdes,PB} makes use of an array to tally the number of times each pair of stars is involved in a potential match. This is a time consuming operation, since it requires all candidate triangles to be processed to allow each one an opportunity to vote. The likely matches are then selected on the basis of probability --- from the pairs that received the highest number of votes. 

By contrast, $OPM$ assumes that highly selective triangles are likely to yield a true match, and if confirmed, the search can be immediately terminated. Therefore, when a potential match is found, the algorithm immediately attempts to verify the relationship using a light-weight (inexpensive) process. If unsuccessful, $OPM$ continues processing more candidates, having expended little effort in screening out the false match. When the preliminary verification is positive, a more robust and relatively expensive verification process is used to comprehensively test the suitability of the match. It is assumed that this process will be executed very few times, most likely only once.

\subsubsection{Preliminary Verification}
\label{sec:PV}
The preliminary verification (PV) process determines the transformation from image to sky coordinates using an astrometric plate solution. It commences with the calculation of standard coordinates ($\xi, \eta$), representing the gnomonic projection of the spherical sky onto the plane of the detector, using the relations
\begin{equation}\label{eq:4}
\xi = \frac{\cos \delta \sin (\alpha - A)}{\sin D \sin \delta + \cos D \cos \delta cos (\alpha - a)}
\end{equation}

\begin{equation}\label{eq:5}
\eta  = \frac{\sin D \cos \delta \cos (\alpha - A) - \cos D \sin \delta} {\sin D \sin \delta + \cos D \cos \delta cos (\alpha - a)} ,
\end{equation} 
where $(\alpha, \delta)$ represent the equatorial coordinates of the catalogue stars and $(A, D)$ is the origin of the coordinates, which is usually taken as the approximate plate center. The standard coordinates are related to the measured coordinates $(x, y)$ of the centroids of the stars on the image using the following relations:
\begin{equation}\label{eq:6}
\xi - \frac{x}{L} = ax + by + c
\end{equation}

\begin{equation}\label{eq:7}
\eta  - \frac{y}{L} = a'x + b'y + c' ,
\end{equation} 
where $a, b, c, a', b', c'$ are the plate constants that describe the translation and rotation necessary to transform between the two coordinate systems, and $L$ is the focal length of the optics, expressed in the same units as $x$ and $y$ \citep{Marsden}. 

The candidate triangle relates three points on the image to three in the reference catalogue, and allows us to write six equations to solve the six unknown plate constants. As a check, we note that $a \sim b'$ and $b \sim -a'$ \citep{Edberg}, assuming that the axes are perpendicular and have the same scale, which should be the case if correct pairings have been selected. If the plate constants differ by more than 2.5\%, a value determined empirically from test images, the candidate pairing is rejected.

\subsubsection{Final Verification}

If the solution appears to be reasonable, a more robust final verification (FV) check is performed. Using the initial plate solution, all $\mathcal{I}$ are transformed to equatorial coordinates and compared to the entire list of $\mathcal{R}$ to find their closest match. An important optimization speeds up this step by avoiding the need to compare all entries. An auxiliary array, containing the indexes into the $\mathcal{R}$ array, was prepared when the $\mathcal{R}$ list was built initially. The auxiliary array was sorted by declination, allowing the $\mathcal{R}$ array to remain sorted by magnitude. Using the auxiliary array, a binary search is performed to locate the starting point within $\mathcal{R}$ where comparisons should commence. The equatorial coordinates of each transformed $\mathcal{I}$ are compared to the catalogue coordinates of all $\mathcal{R}$ that are within $\epsilon$ arcsec. A tolerance of 3$\sigma$ is used, where $\sigma$ is the typical astrometric residual of a full plate solution at this image scale, thus allowing for uncertainties in the initial transformation which is based upon only 3 stars, two of which are closely separated. 

A small angular separation approximation \citep{Meeus} is used to estimate the separation of each pair of stars: 
\begin{equation}\label{eq:8}
s^2 = (\triangle\alpha  \cos \delta )^2 + (\triangle \delta)^2,
\end{equation}
where $s$ is the separation in degrees, $\triangle\alpha$ is their separation in R.A., $\triangle\delta$ is their separation in declination, and $\delta$ is the declination of the target $\mathcal{I}$ (with $\cos \delta$ calculated once \emph{outside} the main loop). The approximation avoids using transcendental functions, which are computationally expensive relative to ordinary floating point operations (addition, multiplication, division). Errors resulting from the approximation are absorbed by the relatively large value of $\epsilon$. The squared separation, $s^2$, is compared to $\epsilon^2$ to avoid a costly $sqrt$ operation. 

Since the $\mathcal{I}$ array is sorted by relative magnitude, the brightest stars are compared first. If multiple $\mathcal{R}$ are found within the matching tolerance, the brightest, unassigned $\mathcal{R}$ is used as the match. This is determined by simply saving the lowest $\mathcal{R}$ index when a match occurs. Since the $\mathcal{R}$ array is sorted by descending magnitude, the saved index represents the brightest $\mathcal{R}$ star. Once an assignment is made, the particular $\mathcal{R}$ is flagged to avoid matching it again. This scheme ensures that the brightest $\mathcal{I}$ are matched to the brightest $\mathcal{R}$ when there are multiple candidates within the matching tolerance, mimicking the decision that a human operator would have made. 

After all assignments have been completed, a new astrometric solution is calculated using all assigned pairs. The process iterates 3 times (this number is user controllable), successively refining the solution at each iteration as more stars are matched. At the end of the process, the final number of matched stars is compared to a predefined limit. If sufficient stars have been identified, the match is deemed to be correct and the search process terminates. In the unlikely event that insufficient stars have been identified, the search process continues with the next candidate.

\subsection{$OPM_B$}

An alternative algorithm, named $OPM_B$, was developed several years ago. I have since learned that it bears some similarity to that described by \citet{Murtagh}. Nevertheless, my approach has some major differences, principally in its use of an early exit strategy and just-in-time approach that avoids calculating quantities until they are required. By postponing various calculations, computational effort is saved in the hope that an early exit will render them unnecessary. 

\begin{figure}[h]
\begin{center}
\includegraphics[scale=0.4, angle=0]{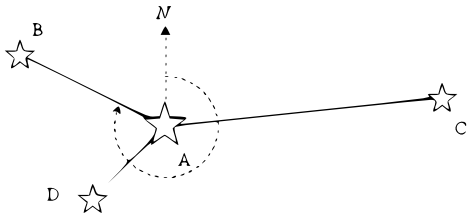}
\caption{\(OPM_B\) constructs shapes of arbitrary complexity using a user-defined number of points.}\label{CSfig}
\end{center}
\end{figure}

$OPM_A$ is dominated by triangle construction costs, particularly for large $n$. $OPM_B$ addresses this problem by reducing the number of shapes to be characterized. It also uses a more restrictive shape definition, which reduces the number of false positives that may occur and results in a successful match being found in nearly constant time, independent of $n$. Instead of matching triangles, an arbitrarily complex geometric shape, made up of a user defined number of points is used (Figure \ref{CSfig}). The shape to be matched is characterized by the relationship of the central star (A) with respect to the other stars (B, C, D, $\ldots$), using their separations and position angles (PA) relative to star A. Angles are measured relative to north (defined as the $-y$ direction as seen from star A), although this is arbitrary.

This definition is similar to that used by \citet{Murtagh}, although his world view describes the relationship of every star to its $n - 1$ neighbors, requiring $O(n^2)$ calculations to describe all points. Furthermore, his world view is calculated for both the $\mathcal{I}$ and $\mathcal{R}$ lists, with the matching process comparing all members of both sets to find a high confidence match, with a resulting computational complexity of $O(n^2)$.  Another point of difference is that Murtagh bins the position-angles into 1$\degree$ increments in order to accommodate rotation of the coordinate systems, with his matching phase requiring the comparison of the world view of set A to 360 versions of set B. Although $OPM_B$ uses a superficially similar shape characterization, the algorithms are quite different. 

In principle, increasing the number of stars used to define the shape adds greater constraints and therefore reduces the number of false matches that may occur. It also allows more points to be used in the initial astrometric solution, leading to a more accurate transformation. In practice, using 3 stars is sufficient because the matching phase is very efficient relative to the shape characterization phase (analogous to triangle construction). The latter dominates the elapsed time of the search, even when false positives are present.

$OPM_B$ processing commences with the lists of the $n$ brightest $\mathcal{I}$ and $\mathcal{R}$, as described in Section 2.1. A sorted list of separations and PAs for each pair of stars in the $\mathcal{R}$ list is constructed. The number of unique pairs, $P$, is given by 
\begin{equation}\label{eq:9}
P = n (n-1) / 2.
\end{equation}
This immediately provides an improvement over $OPM_A$, where two lists of triplets must be prepared instead of one list of pairs ($P \ll T$). 

The search process commences with the selection of the $m$ brightest $\mathcal{I}$, where $m$ is the number of stars used to define the shape to be matched. Each star in the candidate list is assigned a letter, A being the first, B the second, and so on. The separation and PA of each pair, AB, AC, AD, $\ldots$ are calculated. The separations are computed from the focal length of the optical system and physical dimensions of the detector. The PA is calculated relative to the top of the detector, since the absolute rotation relative to the celestial sphere is unknown at this stage. Postponing the same calculations for $\mathcal{I}$ to the search phase avoids the effort of pre-calculating the entire list when only a few values may be required, as is the case when an early exit occurs. This just-in-time approach results in a considerable saving in computational effort.  

The search process attempts to find a match for AB in the list of $\mathcal{R}$ pairs. A binary search is used to quickly identify those pairs with separations within the matching tolerance $\epsilon$. The difference in PA between the image and reference pairs is assumed to be due to rotation of the detector. The $\mathcal{R}$ list is now searched to find candidates for AC, AD, etc. using a binary search on separation and a knowledge of the rotational offset defined by AB with respect to the catalogue value. Candidate pairs with a rotational offset \textgreater 1$\degree$ are rejected. A possible optimization, though not implemented, could reject candidate pairings not matching the absolute orientation of the detector with respect to the sky (when known), thus avoiding the need to determine the rotational offset and permitting incorrect pairs to be rejected immediately.

Once $m$ candidates have been identified, the preliminary verification function is called (see section~\ref{sec:PV}). The $m$ pairings are used to determine an initial astrometric solution which, if acceptable, may result in the final verification process being called. If the solution is rejected, or insufficient candidates are identified, the next candidate within the matching tolerance is selected and the process continues. If the current candidate list cannot be matched, the search process begins again by selecting another set of candidate stars. The process repeats until a successful match is found, or the entire list of candidates is exhausted, in which case we declare that a match could not be found.

\section{Performance}

It is not possible to analytically determine the order of complexity of these algorithms because they do not perform a fixed number of searches. In the best case, a successful match may be found after processing just one candidate. In the worst case, the entire list of candidates may have to be searched. 

In order to investigate whether there is any cause for \emph{optimism}, that is, whether an early match will occur in practice with real data, 10 063 unfiltered, wide-field survey images acquired with a variety of SBIG detectors and focal-lengths were analyzed. Table \ref{T_testimg} lists their characteristics, with the columns describing the focal-length ($f$), number of images, CCD detector, and effective FOV of each set of images. Although this sample of test images was acquired with SBIG detectors, the algorithms are generic in nature and apply equally to all CCD detectors.

Fields were selected from an all-sky survey conducted from a latitude of 35$\degree$S. The deepest, widest fields, located near the galactic equator, contained $\sim3.10^4$ stellar sources to $m_V\sim15$. Images containing moderate defects such as blooming spikes, satellite trails, and thin cirrus were retained in the sample. Images that were heavily obscured by cloud were discarded. In order to test algorithmic robustness under a variety of conditions, approximately 40\% of the images were taken from a photometric survey of bright stars that were strongly defocused to avoid saturation.

\begin{table}[h]
\begin{center}
{\small
\caption{Test Images}\label{T_testimg}
\begin{tabular}{lrcl}
\hline $f$ (mm) & Number & Detector & FOV (deg) \\
\hline 102 & 2498 & ST-6 & 4.8 x 3.6	\\
135     &  211 & ST-6   & 3.6 x 2.8 \\
180$^a$ & 3943 & ST-8XE & 2.9 x 1.9 \\
180     & 1234 & ST-8XE & 4.4 x 2.9 \\
188     & 1102 & ST-8XE & 4.2 x 2.8 \\
200     & 1075 & ST-8XE & 4.0 x 2.6 \\
\hline
\end{tabular}
\medskip\\
$^a$sub-frame\\
}
\end{center}
\end{table}

Elapsed times were measured with the Pentium performance counter (RDTSC instruction), that reports the number of clock cycles that have occurred since the CPU was powered up. Despite its high resolution, precision is limited by unavoidable context switches within the operating system. It is assumed that this effect has been averaged out over the timescale of the test and that each test was affected equally.
 
Separate timers were used to measure the performance of each of the following phases:  triangle (pair) construction, sorting, searching for candidates, preliminary verification, and final verification. In the following discussion, the term \emph{matching} refers to the combined efforts of searching, preliminary verification and final verification.  Although other algorithms do not consider calculation of the transformation (as performed by final verification) to be part of the matching process, it is necessary to include this for $OPM$, since we must be certain that an early exit is warranted. To avoid unfairly penalizing search performance, final verification was configured to use a maximum 100 image stars. 

\begin{figure}[h]
\begin{center}
\includegraphics[scale=0.5, angle=0]{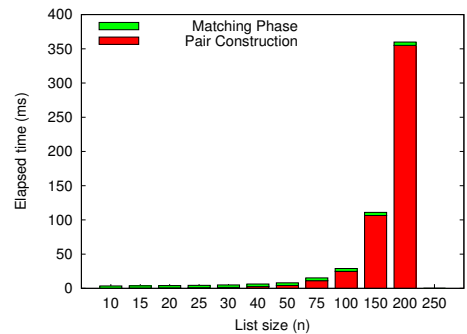}
\caption{$OPM_B$ pair construction and matching phases.}\label{F_opmb_total}
\end{center}
\end{figure}

Results for the two algorithms are summarized in Tables \ref{T_opma} \& \ref{T_opmb}, with the columns describing list size ($n$), total elapsed time, elapsed time for the triangle (pair) construction phase, elapsed time for the matching phase, and the percentage of images successfully matched. Figures \ref{F_opma_total} \& \ref{F_opmb_total} plot the relative construction and matching costs. Figures \ref{F_opma_match} \& \ref{F_opmb_match} show a break-down of the matching phase for each algorithm. Note that the plots use the same vertical scale for easy comparison, and that the abscissa for the $OPM_B$ plots extend to $n = 200$.

\subsection{$OPM_A$ Performance}

\begin{table}[h]
\begin{center}
{\scriptsize
\caption{$OPM_A$ performance}\label{T_opma}
\begin{tabular}{rrrrr}
\hline  $n$ & Total Elapsed & Construct        & Match & Match\\
            & (ms)          & (ms)             & (ms)  & \%   \\
\hline
 10 &   3.99 $\pm$ 0.71 &	  0.23 $\pm$ 0.01 & 3.34 $\pm$ 0.67 &  90.55 \\
 20 &	  6.16 $\pm$ 1.02 &	  2.13 $\pm$ 0.14 & 3.53 $\pm$ 1.00 &  99.92 \\
 30 &	 12.14 $\pm$ 1.75 &	  7.83 $\pm$ 0.19 & 3.60 $\pm$ 1.74 & 100.00 \\
 40 &	 24.43 $\pm$ 1.36 &	 19.74 $\pm$ 0.51 & 3.61 $\pm$ 1.26 & 100.00 \\
 50 &	 46.11 $\pm$ 1.60 &	 40.72 $\pm$ 0.79 & 3.75 $\pm$ 1.38 & 100.00 \\
 60 &	 79.63 $\pm$ 2.16 &	 73.20 $\pm$ 0.92 & 3.96 $\pm$ 1.97 & 100.00 \\
 70 &	128.02 $\pm$ 3.49 &	120.14 $\pm$ 1.38 & 4.30 $\pm$ 3.24 & 100.00 \\
 80 &	194.07 $\pm$ 4.72 &	184.33 $\pm$ 2.85 & 4.66 $\pm$ 3.77 & 100.00 \\
 90 &	280.84 $\pm$ 6.25 &	268.64 $\pm$ 3.73 & 5.25 $\pm$ 5.04 & 100.00 \\
100 &	391.28 $\pm$ 7.84 &	376.01 $\pm$ 4.00 & 6.00 $\pm$ 6.80 & 100.00 \\
\hline
\end{tabular}
\medskip\\
}
\end{center}
\end{table}

\begin{figure}[h]
\begin{center}
\includegraphics[scale=0.5, angle=0]{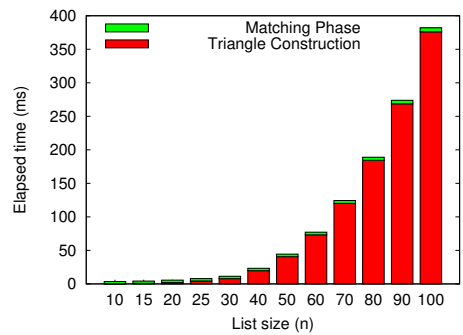}
\caption{$OPM_A$ triangle construction cost relative to the matching phase.}\label{F_opma_total}
\end{center}
\end{figure}

\begin{figure}[h]
\begin{center}
\includegraphics[scale=0.5, angle=0]{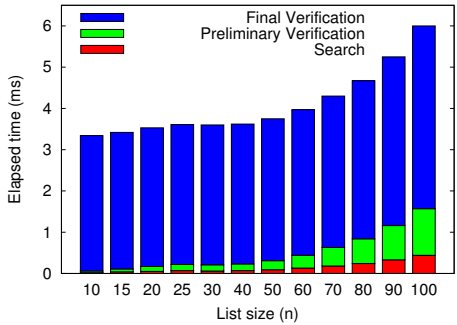}
\caption{$OPM_A$ matching phase only.}\label{F_opma_match}
\end{center}
\end{figure}

The following performance characteristics are observed:  All values of $n \ge$ 30 resulted in a 100\% match rate. Large lists were unnecessary and were in fact detrimental, increasing triangle generation times. Even a small value of $n$ = 30 was sufficient to generate a number of highly selective triangles allowing a match to be found quickly. Values less than 30 did not succeed in matching all images, although somewhat surprisingly, even at $n$ = 10, over 90\% of the images were matched successfully.

Figure \ref{F_opma_total} plots the cost of the matching phase relative to triangle construction. At small $n$, triangle construction costs are negligible and matching dominates (with the majority apportioned to the final verification phase). As $n$ increases, the triangle construction time quickly starts to dominate matching costs, the latter being nearly constant. That triangle construction dominated the total time is in complete contrast to the performance statistics published by \citet{Groth} and \citet{Mars}, where triangle construction was the fast operation and matching dominated. Realizing that triangle construction costs should be similar for all equally optimized algorithms further highlights the effectiveness of the early exit strategy.

The elapsed time in the search and preliminary verification phases is small relative to final verification, confirming that they are suitably light (Figure \ref{F_opma_match}). The cost of final verification could be further reduced by limiting the number of iterations that are performed (3 by default). One could conceivably stop iterating once a sufficient number of stars have been identified, although this optimization was not implemented.

There is very little scatter in total elapsed time, confirming that fast matches, leading to early exits, occur consistently. The median number of candidates processed from the $\mathcal{I}$ triangle list is very small is absolute terms. Less than 1.5\% of $T$ were examined for $n$ = 20, reducing to 0.05\% $T$ for $n$ = 100.

A value of $n$ = 30 appears to be optimal; large enough to produce reliable results and small enough to limit triangle construction and matching costs.  While it is impressive that the entire process can be completed successfully in $\approx$ 12 ms, it is equally remarkable that the cost of searching a much larger list ($n$ = 100) is not prohibitive. This is possible because only a small subset of the triangles is searched instead of processing all combinations.  Nevertheless, large lists offer no practical advantage, particularly when smaller lists are completely reliable. 

\subsection{$OPM_B$ Performance}

$OPM_B$ tests were conducted with $m$ = 3 in order to directly compare the performance to $OPM_A$. It was found that absolute search performance was faster than $OPM_A$. For small values of $n$, both algorithms provide similar performance, due to the relatively large cost of final verification. At $n = 30$, $OPM_B$ is twice as fast as $OPM_A$, due primarily to savings in the construction phase. By $n$ = 100, $OPM_B$ is an order of magnitude faster than $OPM_A$. 

\begin{table}[h]
\begin{center}
{\scriptsize
\caption{$OPM_B$ performance}\label{T_opmb}
 \begin{tabular}{rrrrrr}
\hline $n$ & Total Elapsed & Construct & Match & Match \\
           & (ms)          & (ms)      & (ms)  & \%    \\
\hline 
 10 &   3.56 $\pm$  0.35 &  0.14 $\pm$ 0.03 & 3.32 $\pm$ 0.32 &  87.24 \\
 20 &	  4.09 $\pm$  0.51 &  0.57 $\pm$ 0.03 & 3.42 $\pm$ 0.51 &  99.54 \\
 30 &	  5.04 $\pm$  1.06 &  1.35 $\pm$ 0.05 & 3.58 $\pm$ 1.06 &  99.97 \\
 40 &	  6.39 $\pm$  1.91 &  2.53 $\pm$ 0.16 & 3.75 $\pm$ 1.90 & 100.00 \\
 50 &	  8.19 $\pm$  2.88 &  4.17 $\pm$ 0.13 & 3.91 $\pm$ 2.87 & 100.00 \\
 75 &  15.41 $\pm$  4.45 & 11.19 $\pm$ 0.28 & 4.11 $\pm$ 4.43 & 100.00 \\
100 &	 29.09 $\pm$  5.12 & 24.75 $\pm$ 0.59 & 4.22 $\pm$ 5.06 & 100.00 \\
150 &	111.17 $\pm$  7.69 &106.46 $\pm$ 2.71 & 4.56 $\pm$ 7.18 & 100.00 \\
200 &	360.06 $\pm$ 12.69 &354.98 $\pm$ 8.86 & 4.91 $\pm$ 9.10 & 100.00 \\
\hline
\end{tabular}
\medskip\\
}
\end{center}
\end{table}

\begin{figure}[h]
\begin{center}
\includegraphics[scale=0.5, angle=0]{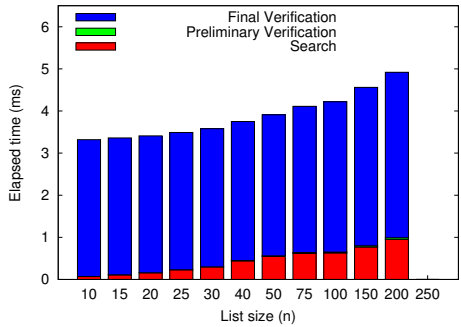}
\caption{$OPM_B$ matching phase only.}\label{F_opmb_match}
\end{center}
\end{figure}

Matching time increased by just 1.6 ms as list size increased from 10 to 200 points. This is attributable to the fact that very few candidates were examined to find a successful match, even for large lists. For $n$ = 100, 60\% of searches were solved using the first candidate list and 90\% of searches were completed by testing $\le$ 10 candidate lists. 

Figure \ref{F_opmb_match} plots the time spent in the sub-phases of matching as a function of $n$. The search time increased by \textless 1 ms between $10 \le n \le 200$, and PV costs were insignificant, due to the use of PA in shape characterization, which removes candidates with incorrect chirality. Final verification accounted for the majority of the time. I expect that further optimizations in the final verification phase might reasonably yield matching times of approximately 1--2 ms.  

\begin{figure}[h]
\begin{center}
\includegraphics[scale=0.5, angle=0]{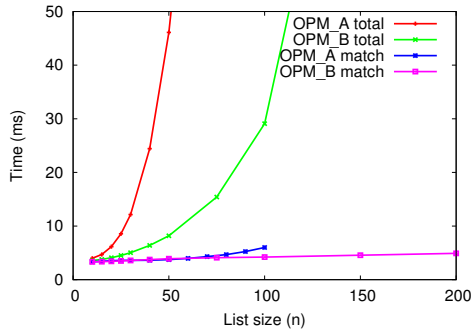}
\caption{Total time vs matching phase.}\label{F_total_vs_match}
\end{center}
\end{figure}

Figure \ref{F_total_vs_match} plots the total elapsed time of the search as a function of $n$, for both $OPM_A$ and $OPM_B$. Also shown is the time spent in each matching phase, highlighting the nearly constant matching time of $OPM_B$. 

\subsection{Relative Performance}

A number of authors have provided indicative performance measurements for their respective implementations. Unfortunately, absolute timings are difficult to compare because they are quoted for different values of $n$, statistics are not provided for all phases, and differences in machine architecture and processor speed play a significant role in determining the overall performance. Nevertheless, it is possible to make some general observations by comparing recent results produced on a similar CPU. 

Most recently, \citet{PB} demonstrated a mean elapsed time of $\sim$100ms to process a full-triangulation of 35 sources using their \textbf{grmatch} task, which implements a voting algorithm (2.0 GHz 64-bit AMD Opteron CPU). The time quoted for \textbf{grmatch} excluded iterative calculation and refinement of the transformation coefficients. Table \ref{T_opmb} shows that $OPM_B$ completes the same task in $\sim$6 ms, including the extra work of final verification. Even allowing for an $\sim$20\% difference in processor speed, it is clear that early exits are extremely beneficial, with the performance differential expected to widen as $n$ increases.  

\subsection{Ill-conditioned Searches}

The preceding tests were performed on wide-field images where pointing errors were small relative to the size of the FOV. Thus, there was nearly a 100\% overlap between $\mathcal{I}$ and $\mathcal{R}$. We now consider the performance of $OPM_B$ under non-optimal conditions. 

\begin{figure}[h]
\begin{center}
\includegraphics[scale=0.6, angle=0]{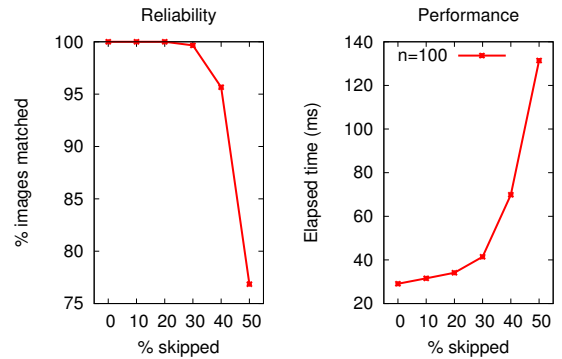}
\caption{$OPM_B$ reliability and performance when skipping the brightest $\mathcal{I}$ stars.}\label{F_opmb_csskip}
\end{center}
\end{figure}

Figure \ref{F_opmb_csskip} plots the match rate and elapsed time of 10 063 searches ($n = 100$) when the brightest stars have been omitted from $\mathcal{I}$, as might be the case if they were saturated or a significant passband disparity exists. A 100\% match rate was maintained even when skipping 20\% of $\mathcal{I}$, dropping slightly to 99.7\% at 30\% of $\mathcal{I}$. The elapsed time was only marginally affected for values up to 30\%, but did increase markedly when a significant fraction of stars were skipped because the mismatched lists reduced the likelihood of an early exit being taken. Nevertheless, skipping (an unrealistic) 30\% of stars did not significantly affect reliability or performance. 

\begin{figure}[h]
\begin{center}
\includegraphics[scale=0.6, angle=0]{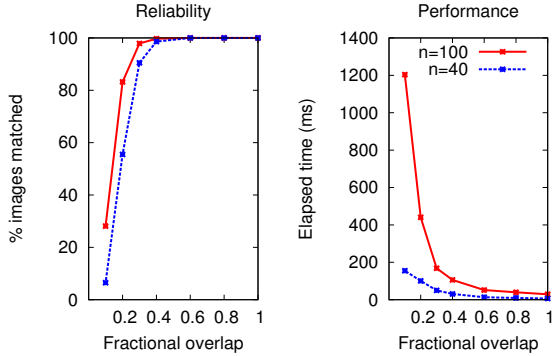}
\caption{$OPM_B$ reliability and performance for partially overlapped fields.}\label{F_opmb_csfrac}
\end{center}
\end{figure}

Figure \ref{F_opmb_csfrac} plots $OPM_B$ performance for partially overlapping fields. Scenarios where $\mathcal{I}$ and $\mathcal{R}$ are not aligned are more typical of narrow-field images, where pointing errors may be a significant fraction of the FOV. Curves are plotted for two values of $n$. A value of $n=100$ was slightly more reliable than $n=40$, but the latter performed far better as the degree of overlap decreased. Under these conditions, smaller values of $n$ are favored to avoid long search times when the chance of finding a successful match is small. If the coordinates of the field center are unknown, an iterative (perhaps spiral) search should use a small $n$ to reduce the elapsed time of any unsuccessful (exhaustive) searches. 

\begin{figure}[h]
\begin{center}
\includegraphics[scale=0.6, angle=0]{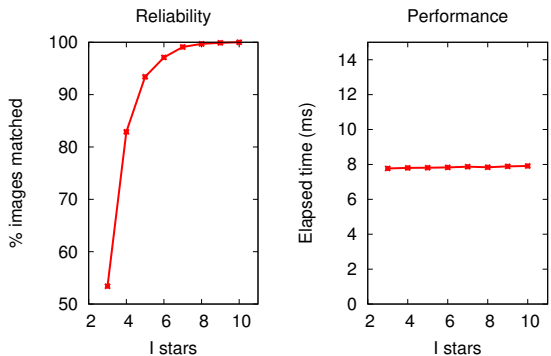}
\caption{$OPM_B$ reliability and performance for fields containing very few stars ($3 \le \mathcal{I} \le 10$ and $\mathcal{R}=50$).}\label{F_opmb_cssmall}
\end{center}
\end{figure}

Extremely narrow fields of view were simulated by conducting tests using very few $\mathcal{I}$ stars. Figure \ref{F_opmb_cssmall} plots reliability and performance when constraining $3 \le \mathcal{I} \le 10$. Only the well-focused subset of 6120 images was used, so that spurious stellar detections from blended defocused objects would not be included within $\mathcal{I}$, which would otherwise skew results. The size of the reference list was set to $\mathcal{R}=50$ to reduce the chance of passband disparities producing non-overlapping lists, which is more likely when both $\mathcal{I}$ \emph{and} $\mathcal{R}$ are small. The plot shows that as few as 5 stars were sufficient to successfully match 93.4\% of cases, rising to 100\% at $\mathcal{I}=10$. This is in contrast to the results shown in Table \ref{T_opmb} for $n \le 30$, which were less reliable because \emph{both} lists were small, resulting in a reduced match rate. The combination of $\mathcal{I}=10$, $\mathcal{R}=50$ provides both high reliability and good performance, with searches completing in $\sim 8$ ms.

\section{Summary}

Two new techniques for matching two-dimensional coordinate lists in nearly constant time have been presented. The matching phase of $OPM_B$ is nearly $O(1)$, being independent of list size. These algorithms have a significant performance advantage over previous techniques, at a slight loss in generality, caused by the requirement that the approximate focal length of the optical system is known \emph{a priori}. This requirement permits the determination of the image scale from the physical dimensions of the detector, allowing $OPM$ algorithms to directly compare a subset of triangles (or shapes) to their counterparts derived from a reference catalogue, without having to process the entire set, as is the case when the scale is unknown. By employing early exit strategies, postponing work until absolutely necessary, testing candidates in the order most likely to yield success, and combining these with and an efficient mechanism for rejecting false positives, a highly efficient search, in nearly constant time is possible.

Small uncertainties in the focal length, such as caused by temperature related changes, are accommodated by selecting an appropriate matching tolerance. The actual focal-length is determined and reported as part of the astrometric solution.

The $OPM$ algorithms are particularly suited to processing large lists or in situations where pattern matching must be performed as quickly as possible. The performance of these algorithms makes it practical to search thousands of fields very quickly, if for example, the coordinates of the field center were unknown. Similarly, when only an approximate focal-length is known, it is perfectly reasonable to attempt to iteratively match the field using a range of focal-lengths. 

\section*{Acknowledgments} 
I would like to thank Tim Bedding and Laszlo Kiss, who kindly reviewed this manuscript and offered many useful suggestions to improve its quality. Similarly, I thank and acknowledge the anonymous referee who suggested a number of valuable avenues to explore.


\end{document}